\documentclass[10pt,journal]{IEEEtran}
\usepackage{times}
\usepackage[cmex10]{amsmath}
\usepackage{amssymb}
\usepackage{epsfig,verbatim}
\usepackage{algorithm,algpseudocode} 
\usepackage{hhline}
\usepackage{algpseudocode}
\usepackage{booktabs}
\usepackage{multirow}
\usepackage{graphicx, subfigure}
\usepackage{cite}

\long\def\comment#1{}

\newfont{\bbb}{msbm10 scaled 700}

\newfont{\bb}{msbm10 scaled 1100}

\newcommand{\RR}{\mbox{\bb R}}

\newcommand{\rv}{{\bf r}}

\newcommand{\wv}{{\bf w}}

\newcommand{\xv}{{\bf x}}

\newcommand{\zv}{{\bf z}}

\newcommand{\Id}{{\bf I}}

\newcommand{\Ac}{{\cal A}}

\newcommand{\Cc}{{\cal C}}
\newcommand{\Dc}{{\cal D}}
\newcommand{\Ec}{{\cal E}}

\newcommand{\Mc}{{\cal M}}
\newcommand{\Nc}{{\cal N}}

\newcommand{\Pc}{{\cal P}}
\newcommand{\Qc}{{\cal Q}}
\newcommand{\Rc}{{\cal R}}
\newcommand{\Sc}{{\cal S}}
\newcommand{\Tc}{{\cal T}}
\newcommand{\Uc}{{\cal U}}

\newcommand{\epsilonv}{\hbox{\boldmath$\epsilon$}}

\newcommand{\muv}{\hbox{\boldmath$\mu$}}

\newcommand{\thetav}{\hbox{\boldmath$\theta$}}

\newcommand{\SNR}{{\sf SNR}}

\newcommand{\eqdef}{\stackrel{\Delta}{=}}

\newcommand{\transp}{{\sf T}}

\usepackage{xcolor}

\newcommand{\argmax}{\operatornamewithlimits{argmax}}
\newcommand{\argmin}{\operatornamewithlimits{argmin}}

\title{A Supervised-Learning Detector for Multihop Distributed Reception Systems}

\author{
\IEEEauthorblockN{
              Seonho Kim and Song-Nam Hong\\}
\thanks{ Copyright (c) 2015 IEEE. Personal use of this material is permitted. However, permission to use this material for any other purposes must be obtained from the IEEE by sending a request to pubs-permissions@ieee.org.

The authors are with the Department of Electrical Engineering, Ajou University, Suwon, Korea (e-mail: \{kimsh1005, snhong\}@ajou.ac.kr). 

This work was supported by Samsung Research Funding $\&$ Incubation Center of Samsung Electronics under Project Number SRFC-IT1702-00.
}

}

\begin{document}

\maketitle

\date{}


\begin{abstract}
We consider a multihop distributed uplink reception system in which $K$ users transmit independent messages to one data center of $N_{\rm r} \geq K$ receive antennas, with the aid of multihop intermediate relays. In particular, each antenna of the data center is equipped with  one-bit analog-to-digital converts (ADCs) for the sake of power-efficiency. In this system, it is extremely challenging to develop a low-complexity detector due to the non-linearity of an end-to-end channel transfer function (created by relays' operations and one-bit ADCs). Furthermore, there is no efficient way to estimate such complex function with a limited number of training data. Motivated by this, we propose a supervised-learning (SL) detector by introducing a novel Bernoulli-like model in which training data is directly used to design a detector rather than estimating a channel transfer function. It is shown that the proposed SL detector outperforms the existing SL detectors based on Gaussian model for one-bit quantized (binary observation) systems. Furthermore, we significantly reduce the complexity of the proposed SL detector using the fast kNN algorithm. 
Simulation results demonstrate that the proposed SL detector can yield an attractive performance with a significantly lower complexity.
\end{abstract}

\section{Introduction}

A distributed uplink reception system is a special case of a  multi-source single-destination multihop relay network where multiple sources send independent messages to one destination of a large number of antennas with the help of multihop intermediate relays. In this system, numerous information-theoretical approaches have been proposed in  \cite{Avestimehr,Lim,Park,Park1}, with the assumption that the destination  perfectly knows all channel transfer functions (or at least end-to-end channel transfer function). A quantized-remap-and-forward (QMF) (extended in \cite{Lim} where it is referred to as noisy network coding (NNC)) was presented in \cite{Avestimehr}, which achieves the best-known performance. However, it is not practical as joint typical detector at the destination is prohibitive and  the assumption of perfect channel state information is unrealistic. A more practical approach based on lattice code, named compute-and-forward (CoF), was presented in \cite{Bobak,Hong,Hong-TVT}, which can significantly decrease the detection complexity, by converting the non-linear end-to-end channel transfer function into a linear one. However, its performance is not satisfactory for multihop relay networks with  realistic channels (e.g., Rayleigh fading), due to a severe  non-integer penalty \cite{Hong}. Therefore, it is still an open problem to develop a practical detection and channel estimation methods for a multihop communication system.

In a distributed uplink reception system, the use of a large number of receive antennas at data center is necessary to support multiple sources simultaneously. Unfortunately, it can highly increase the hardware cost and  the radio-frequency (RF) circuit consumption \cite{yang2013total}. Especially, a high-resolution analog-to-digital converter (ADC)  is most problematic as the power consumption of an ADC is scaled exponentially with the number of quantization bits and linearly with the baseband bandwidth~\cite{ADC, modelling}. To overcome this, the use of low-resolution ADCs (e.g., 1 $\sim$ 3 bits) has received increasing attention for a large-scale multiple-input-multiple-output (MIMO) system  \cite{pohang, Hong_J, Hong-Soft, Jeon}. The one-bit ADC is particularly attractive as it does not need an automatic gain controller~\cite{monobit}. In this sense, we consider a multihop distributed uplink reception system in which each receive antenna of the data center is equipped with one-bit ADCs. For this system, an end-to-end channel transfer function  between $K$ users and the data center is highly non-linear. Thus, it is extremely challenging to estimate such function with a limited number of one-bit quantized pilot signals. This motivates us to consider a data-driven supervised-learning (SL) detector in which the pilot signals (or training data) are exploited to directly learn a MIMO detector rather than estimating a complicated non-linear channel transfer function.

Very recently, SL detectors have been developed in \cite{pohang} for MIMO systems with one-bit ADCs. It is remarkable that these methods are developed by assuming that data is generated from a Gaussian distribution. Although it is widely used, this model might not be suitable for binary data (e.g., one-bit quantized observations). In this paper, we propose a novel Bernoulli-like model which can be more suitable for binary random outputs. It is verified by showing that the proposed SL detector outperforms the existing SL detectors in \cite{pohang}. Despite its superior performance, the complexity of the proposed SL detector (also, the existing SL detectors in \cite{pohang}) is problematic as a search-space grows exponentially with the number of users $K$. This is the major drawback to be used in practice. We address this problem by presenting a low-complexity SL (LSL) detector using  the  {\em fast kNN} algorithm. The fast kNN algorithm, which can find a closest point in Hamming space fastly using an efficient data structure, enables to efficiently remove unnecessary elements in the search-space according to a given current observation. Thus, the LSL detector can perform over the significantly reduced search-space. Simulation results demonstrate that the proposed LSL detector can yield an attractive performance with a practically manageable complexity.


This paper is organized as follows. In Section~\ref{system}, we describe a multihop distributed uplink reception system. In Section \ref{detector}, we briefly review the various existing SL detectors. In Section~\ref{propose}, we propose a novel SL detector based on a Bernoulli-like model. In Section \ref{lowcomde}, we significantly reduce the complexity of the proposed SL detector by efficiently using the fast kNN algorithm. Section \ref{simulation} provides the simulation results to verify the superiority of the proposed SL detector. Finally, conclusion is provided in  Section \ref{conclusion}.

\section{System model}\label{system}

We consider a multihop distributed uplink reception system where $K$ sources transmit independent messages to one data center with the help of intermediate relays. In particular, the data center is equipped with $N_{\rm r} \geq K$ receive antennas with one-bit ADCs. Let $w_k\in \{{0,..,m-1}\}$ denote the source $k$'s message for $k\in\{1,...K\}$, each of which contains the $\log{m}$ information bits. We also denote $m$-ary constellation set by $S=\{s_0,...,s_{m-1}\}$ with power constraint $\frac{1}{m}\sum_{i=0}^{m-1}|s_i|^2=P_{\rm t}$.
Let $\mbox{sign}(\cdot): \RR \rightarrow \{-1,1\}$ represent the one-bit ADC quantizer function with $\mbox{sign}(u)=1$ if $u\geq 0$ and $\mbox{sign}(u)=-1$, otherwise. Then, the transmitted symbol of the source $k$, ${\tilde x}_k$, is obtained by a modulation function $f:{W}\rightarrow \Sc$ as $\tilde{x}_k=f(w_k) \in \Sc$. Then, by converting a complex-valued scalar into an equivalent real-valued vector, the data center observes
\begin{equation}\label{eq:c_trans}
\rv=\mbox{sign}(\Phi(\tilde{\xv})+\tilde{\zv}) \in \{-1,1\}^N,
\end{equation} where $N=2N_{\rm r}$ and $\Phi(\cdot)$ represents a complex {\em non-linear} function (called end-to-end channel transfer function). Also,  $\tilde{\zv}=[\tilde{z}_1,\ldots,\tilde{z}_{N}]\in\mathbb{R}^{N}$ denotes the noise vector whose elements are independent and identically distributed as circularly symmetric complex gaussian random variables with zero-mean and variance $\sigma_z^2$, i.e., ${\tilde z}_i \sim \Cc\Nc(0,\sigma_z^2)$. 

It is remarkable that $\Phi(\cdot)$ can capture all the intermediate relays' operations and all the local wireless channels in the network.  Although the proposed method in this paper can be applied to any relay's operation and local channel model, we assume that in our simulations, each relay with a single antenna performs an amplify-and-forward (AF) and each local channel is assumed as Rayleigh fading. Also, for the simplicity, it is assumed that each relay has the same power constraint with the sources as $P_{\rm t}$ and all the additive noises at the receivers in the network are circularly symmetric complex gaussian random variables with zero-mean and variance $\sigma_z^2$. We define the signal-to-noise ratios (SNRs) as 
 $\SNR =P_{\rm t}/\sigma_z^2$.

\begin{figure}
\centerline{\includegraphics[width=8cm]{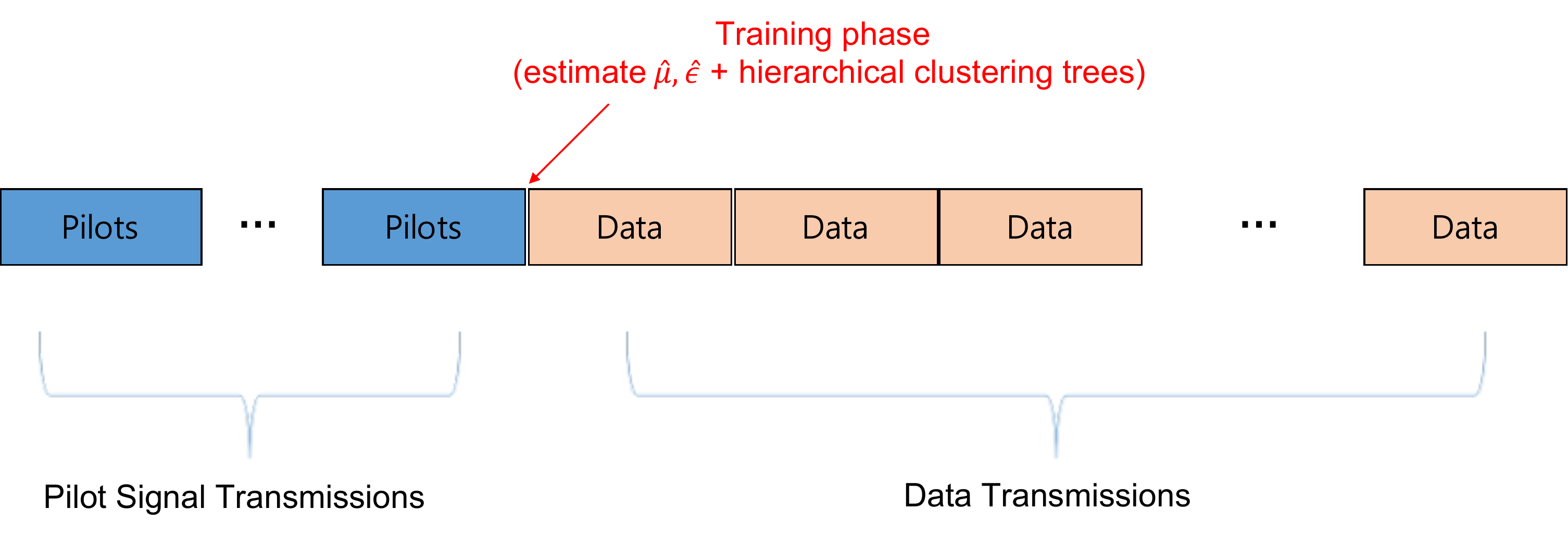}}
\caption{Illustration of the training and data transmission phases within a coherence time.}
\label{phase}
\end{figure}

The proposed communication framework consists of training and data transmission phases  (see Fig.~\ref{phase}). 
 Note that during these phases, a wireless channel is assumed to be fixed.
\begin{itemize}
\item {\bf Training phase:} In this phase,  $K$ sources transmits ``known" sequences (i.e., pilot signals) so that the data center can learn a non-linear function $\Phi(\cdot)$. With machine-learning perspective, the data center collects the data and the corresponding labels. Let $\Mc=\{0,\ldots,m-1\}^{K}$ denote the set of all possible messages of the $K$ sources. For each class $c \in \Mc$, the $K$ sources  transmit $T$ pilot signals $\tilde{\xv}^c_{i}$ for $i=1,...,T$.  From (\ref{eq:c_trans}), the data center can collect the {\em labelled} data set 
\begin{equation}
\Dc=\left\{\tilde{\rv}^{c}_{i} \in \{-1,1\}^N: c \in \Mc, i=1,...,T\right\}.
\end{equation}

\item {\bf Data transmission phase:} Given the $\Dc$ and a new observation $\rv$, the data center detects the class of $\rv$ (i.e., users' messages $\hat{\wv}=({\hat w}_1,...,\hat{w}_K)$) as 
\begin{equation}
\Psi_{\Dc}(\rv) = c \in \Mc,
\end{equation} 
which is what we will propose in this paper.
\end{itemize}

\section{SL Detectors for Binary Data}\label{detector}

In machine-learning perspective, the above detection problem (a.k.a., the supervised-learning problem) can be categorized into two approaches \cite{murphy} as {\em non-parametric} and {\em parametric} learnings.  A non-parametric learning does not require a  priori knowledge on data set $\Dc$ (e.g., a distribution of data) such as  $k$-nearest neighbor (kNN), decision tree, and support vector machine (SVM).  Whereas, in parametric learnings as logistic regression, naive bayes, and neural networks, data is assumed to be generated from a probabilistic model with some parameters (e.g., Gaussian model).  Then, they are optimized from the given data set $\Dc$. Therefore, it is very important to choose a proper probabilistic model based on a priori knowledge (or domain knowledge) on the data set $\Dc$.

We briefly review the existing (parametric or non-parametric) SL detectors. It is noticeable that they can be immediately applied to a distributed reception system since the SL detector do not rely on system models.

\begin{itemize}

\item {\bf Non-parametric learning:} In \cite{pohang}, empirical maximum-likelihood detector (eMLD) and minimum mean distance detector (MMD) have been presented. The eMLD can be viewed as kNN classifier where the $k$ nearest data points from a new observation (or received signal) are identified and then, the majority voting is performed to find a class (e.g., users' messages). Also, the MMD is the special case of eMLD with $k=1$ for the purpose of low-complexity.

\item {\bf Parametric learning:}  In this approach, it is most important to seek a proper probabilistic model for a given data set $\Dc$. As in \cite{pohang,murphy}, a Gaussian model is widely used where the data $\rv \in \Dc$ is assumed to be generated from  the probability distribution $P(\rv| c,\thetav_{\rm{c}})=\Nc \left(\muv_c,\Sigma_c \right)$. Here, $c\in \Mc$ denotes the class (or message) of the $K$ sources and $\thetav_{c}$ represents the parameter vector for the class $c$. Using the given $\Dc=\{\tilde{\rv}_{t}^c: t=1,...,T\}$, we can optimize $\thetav_{c}=(\hat{\muv}_{c}, \hat{\Sigma}_{c})$ via maximum likelihood (ML) estimation as
\begin{align}
\hat{\muv}_c&=\frac{1}{T}\sum_{t=1}^{T} {\tilde{\rv}^c_t} \label{mean} \\
\hat{\Sigma}_c&=\frac{1}{T}\sum_{t=1}^{T}(\tilde{\rv}^c_t-\hat{\muv}_c)(\tilde{\rv}^c_t- \hat{\muv}_c)^{\transp}, \label{variance}
\end{align} where $\hat{\muv}_{c}$ and $\hat{\Sigma}_{c}$ represent the mean and the covariance of the training data associated with the class $c$, respectively. When the training data is not sufficient, the covariance matrix tends to be rank-deficient and ill-conditioned. This problem can be resolved by shrinkage estimator \cite{shrinkage}. Given the $\hat{\thetav}_{c} = (\hat{\muv}_{c}, \hat{\Sigma}_{c})$, the optimal ML detector is derived as
\begin{equation}\label{NCC}
\Psi_{\Dc}(\rv)=\argmin_{c \in \Mc}\; (\rv-\hat{\muv}_c)^{\transp}\hat{\Sigma}_c^{-1}(\rv-\hat{\muv}_c).
\end{equation} In particular, the distance measure in the above is referred to as Mahalanobis distance, and the inverse matrix of $\hat{\Sigma}_{c}$ in (\ref{variance}) is called a precision matrix. When $\Sigma_c = \Id$ for all $c$, as a special case, the resulting detector is equivalent to the Minimum-Centered-Distance (MCD) detector proposed in \cite{pohang}.
\end{itemize}
It was shown in \cite{pohang} that, among the above SL detectors, MCD and eMLD detectors show the best performances. Since the complexity of eMLD is higher than MCD, the latter was highly recommended. However, one can argue that Gaussian model in (\ref{NCC}) might not be suitable to model the distribution of binary data $\rv\in\{1,-1\}^{N}$. This motivates us to propose a SL detector using a novel Bernoulli-like probabilistic model (see Section~\ref{propose}).

\section{The Proposed (Parametric) SL Detector}\label{propose}

We propose a novel SL detector based on a Bernoulli-like model, where data is assumed to be generated from the following probability distribution:
\begin{equation}\label{eq:B-model}
P(\rv |c,\thetav_{\rm{c}})=\prod_{i=1}^{N} \epsilon_{\rm c,i}^{{\bf 1}_{\{r_i \neq \mu_{c,i} \}}}(1-\epsilon_{c,i})^{{\bf 1}_{\{r_i=\mu_{c,i}\}}},
\end{equation} where $\thetav_c=(\muv_c, \epsilonv_c$) for $c\in \Mc$, $\epsilon_{c,i} < 0.5$ for all $i$, and ${\bf 1}_{\{\Ac\}}$ represents an indicator function with ${\bf 1}_{\{\Ac\}}=1$ if $\Ac$ is true, and ${\bf 1}_{\{\Ac\}}=0$, otherwise. Given the training data for the class $c$ (e.g., $\{\tilde{\rv}_{t}^c: t=1,...,T\}$), the parameter vector $\thetav_c$ is optimized using ML estimation as
\begin{equation}\label{object}
(\hat{\muv}_{c}, \hat{\epsilonv}_c)=\argmax_{(\muv_c,\epsilonv_c)}\prod_{t=1}^{T}P(\tilde{\rv}_t^c| \muv_c, \epsilonv_c).
\end{equation}  By plugging (\ref{eq:B-model}) into (\ref{object}), the optimal parameters are obtained by taking the solutions of
\begin{equation*}\label{eq:opt2}
(\hat{\muv}_{c}, \hat{\epsilonv}_c)=\argmax_{(\muv_c,\epsilonv_c)}\prod_{i=1}^{N}\prod_{t=1}^T \epsilon_{c,i}^{{\bf 1}_{\{\tilde{r}_{t,i}^c\neq \mu_{c,i}\}}} (1-\epsilon_{c,i})^{{\bf 1}_{\{\tilde{r}_{t,i}^c=\mu_{c,i}\}}}.
\end{equation*} 
For any $\epsilon_{c,i}<0.5$, we can see that the above objective function is maximized by taking 
\begin{equation}\label{mu}
\hat{\mu}_{c,i}=\mbox{sign}\left(\sum_{t=1}^{T} \tilde{r}_{t,i}^c\right) \mbox{ for } i=1,...,N,
\end{equation} independently from the choices of $\epsilon_{c,i}$'s. We let
\begin{equation}
N_{\rm d}= \sum_{t=1}^{T} {{\bf 1}_{\{\tilde{r}_{k,i}^c\neq \hat{\mu}_{c,i}\}}} \mbox{ and } N_{\rm s}= \sum_{t=1}^{T} {{\bf 1}_{\{\tilde{r}_{k,i}^c= \hat{\mu}_{c,i}\}}}.
\end{equation} Then, we can find an optimal $\epsilon_{c,i}$ independently from the other $\epsilon_{c,j}$'s with $i\neq j$ by taking the solution of $\argmax_{\epsilon_{c,i}} \epsilon_{c,i}^{N_d}(1-\epsilon_{c,i})^{N_s}$.
Taking $\frac{\partial (\epsilon_{c,i}^{N_d}(1-\epsilon_{c,i})^{N_s}) }{\partial \epsilon_{c,i}}=0$, the optimal $\epsilon_{c,i}$ is obtained as
\begin{equation}\label{eq:opt_epsilon}
\hat{\epsilon}_{c,j}=\frac{1}{T}\sum_{t=1}^T {\bf 1}_{\{\hat{\mu}_{c,j}\neq \tilde{r}_{j,i}^c\}}.
\end{equation} With the parameter vector $\hat{\thetav}_{c} = (\hat{\muv}_{c}, \hat{\epsilonv}_{c})$ in (\ref{mu}) and (\ref{eq:opt_epsilon}), the optimal ML estimator (i.e., the proposed SL detector) is derived as
\begin{align}\label{bNCC}
&\Psi_{\Dc}(\rv)=\argmin_{c \in \Mc} \; (\rv - \hat{\muv}_c)^{\transp}\mbox{\bf diag}\left[-\log{\hat{\epsilon}_{\rm{c,i}}}\right] (\rv - \hat{\muv}_c),
\end{align} where $\mbox{\bf diag}[d_i]$ denotes the diagonal matrix with the $i$-th diagonal element $d_i$ and its dimension is easily obtained from the context.

\section{The Proposed LSL Detector}\label{lowcomde}

In the proposed SL detector in (\ref{bNCC}), the computational complexity is expensive as the size of search-space (e.g., $|\Mc|=m^K$) grows exponentially with $K$. To address this problem, we present a low-complexity SL (LSL) detector which is performed over the {\em reduced} search-space. The major contribution of this section is to build the reduced search-space by efficiently removing unnecessary candidates from the $\Mc$ according to a current observation $\rv$.

The proposed method to yield the reduced search-space can be outlined as follows (see Fig~\ref{phase}):

{\bf Training phase:} From the training data $\Dc=\{\rv_{t}^c: t=1,...,T, c\in \Mc\}$, the parameters for the proposed SL detector are obtained from (\ref{mu}) and (\ref{eq:opt_epsilon}) as $\hat{\Uc}=\{\hat{\muv}_c: c\in \Mc\}$ and $\hat{\Ec}=\{\hat{\epsilonv_{c}}: c \in \Mc\}$. Then, $\hat{\Uc}$ is decomposed using $k$-medoids clustering  in \cite{flann}, yielding a hierarchical clustering tree (see Algorithm 1). This algorithm starts with all the elements in $\Dc$ and decomposes them into $J$ clusters, where $J$ is a parameter of the algorithm and called branching factor. The clusters are constructed by selecting $J$ elements randomly as cluster centroids and then by assigning other elements to one of the clusters with the closest centroid. The algorithm is repeated recursively until the number of elements in each cluster is below the maximum leaf size $J$, where in this case, that node becomes a leaf node. In addition, Algorithm 1 is performed over $W$ times to construct the $W$ trees having possibly different decomposition structures, denoted by $\{ {\cal T}_1,\ldots,{\cal T}_W\}$. The use of the multiple trees can improve the quality of the resulting reduced search space.

{\bf Data transmission phase:} Given a current observation $\rv$,  the search algorithm begins with traversing multiple trees in parallel. Note that $W$ multiple trees share a single priority queue ($\Qc$) where the nodes in the priority queue are arranged in the shortest Hamming distance order, with respect to the current observation $\rv$. Then, it can efficiently produce the reduced search-space $\Sc(\rv)\subseteq \Mc$ which only contains the nearest $\hat{\muv}_{c}$'s to the $\rv$. The detailed procedures are given in Algorithm 2. Then, the proposed SL detector is performed as
 \begin{equation}\label{lowcom}
\Psi_{\Dc}(\rv)=\argmin_{c \in \Sc(\rv)} \; (\rv - \hat{\muv}_c)^{\transp}\mbox{diag}\left[-\log{\hat{\epsilon}_{\rm{c,i}}}\right](\rv - \hat{\muv}_c).
\end{equation}

\begin{algorithm}\label{algoritm1}
\caption{Hierarchical clustering tree $h(\hat{\Uc},J)$}\label{alg:euclid}
\textbf{Input:} $\hat{\Uc}=\{\hat{\muv}_1,...,\hat{\muv}_{|\Mc|}\}$\\
\textbf{Output:} hierarchical clustering tree ${\cal T}$\\
\textbf{Parameter:} $J$ (maximum leaf size and branching factor)
\begin{algorithmic}
\If {$\textit{$|\hat{\Uc}| < J$}$}
\State $\textit{create leaf node with the elements in $\hat{\Uc}$}$
\Else 
\State $\Pc \gets \textit{select J elements randomly from $\hat{\Uc}$ (centroids)}$
\State $\Cc \gets$ cluster the elements in $\hat{\Uc}$ with the centroids $\Pc$
\For \mbox{ each cluster} {$\Cc_i \in \Cc=\{\Cc_1,...,\Cc_J\}$}
 \State \textit{create non-leaf node with centroid $\Pc_i$} 
 \State \textit{recursively apply the algorithm $h(\Cc_i, J)$ (with the updated $\Cc_i$)}
\EndFor
\EndIf
\end{algorithmic}
\end{algorithm}

From now on, we will analyze the computational complexity of the proposed LSL detector which consists of construction and search complexities. Note that the construction complexity is taken only once during each coherence time and thus, it can be negligible when the coherence time is sufficiently large (i.e., a channel is slowly changed). Recall that $N$ denotes the observation dimensionality (e.g., $N=2N_r$) and  $m^K$ denote the number of all possible messages of $K$ users.

{\bf Construction complexity:}  On the construction of $k$-medoids hierarchical trees, the complexity of distance-computation is equal to $\mathcal{O}(m^K N)$ with respect to a node selected as cluster centroid. Given the branch factor $J$, it constructs $J$ distinctive clusters at each tree level and thus, the corresponding complexity is equal to $\mathcal{O}(m^K NJ)$. Assuming that $W$ multiple trees have balanced structures, the height of the tree will be 
${K\log_{J}{m}}$. Then, the overall complexity for the forest construction is equal to $\mathcal{O}({m^K}NJW{K\log_{J}{m}})$.

\begin{algorithm}\label{algoritm}
\caption{Searching parallel hierarchical clustering trees}\label{alg:euclid}
\textbf{Input:} hierarchical clustering trees $\{{\cal T}_i: i=1,...,W\}$ and a new observation $\rv$ \\
\textbf{Output:} $\Sc(\rv)$ (reduced search space associated with $\rv$)\\ 
\textbf{Parameter:} $L_{\rm max}$ (the desired size of a reduced search space, e.g., $|\Sc(\rv)|=L_{\rm max}$)
\begin{algorithmic}[1]
\State $\ell \gets \text{0}  \text{  ($\ell$ = }\textit{number of points $\hat{\muv}_c$ searched} \text{)}$
\State $\Qc \gets \textit{empty priority queue}$ 
\State $\Rc \gets \textit{empty priority queue}$
\For{ each tree $\Tc_i$ }
 \State $\textit{call} \text{ TraverseTree($\Tc_i$, $\Qc$, $\Rc$)} $
\EndFor
\While{ $|\Qc| \neq 0$ \textit{and} $\ell < L_{\rm max}$} 
\State $j \gets \textit{top index of $\Qc$}$
\State call \text{ TraverseTree($j,\Qc,\Rc$)}  
\EndWhile \\
\Return \textit{K top points from $\Rc$}
\end{algorithmic}
\begin{algorithmic}[1]
\Procedure{TraverseTree}{$j,\Qc,\Rc$}
\If {$\textit{node j is a leaf node}$}
\State $\Sc \eqdef \{\mbox{all the elements in the leaf node } j\}$ 
\State $\Rc = \Rc \cup \Sc$ and $\ell \gets \ell + {|\Sc|}$
\Else 
\State $\Cc \gets \textit{child nodes of node $j$}$
\State $i \gets \textit{closest node of $\Cc$ to the \rv}$
\State $\Cc_p \gets \Cc \setminus \{i\}$ 
\State \textit{add all nodes in $\Cc_p$ to $\Qc$}
\State {call} \text{ TraverseTree($i$, $\Qc$, $\Rc$)}
\EndIf
\EndProcedure
\end{algorithmic}
\end{algorithm}

{\bf Search complexity:} First, it starts with traversing the forest $\{{\cal T}_1,\ldots,{\cal T}_{W}\}$ simultaneously. Each node computes the distances with the $J$ child nodes to find the closest node at each level. This computation repeats until it reaches to a leaf node at level ${K\log_{J}{m}}$ for each tree ${\cal T}_i$. The corresponding complexity is  $\mathcal{O}(WJN{K\log_{J}{m}})$. According to the Algorithm 2, it stops after examining $L_{max}$ elements. Except the first step, the remaining steps start with a node popped out from priority queue, which is likely to be located at a level between root and leaf.  Assuming that every search returns $J$ elements in a reached leaf node and the starting point is the root, the corresponding complexity is equal to $\mathcal{O}(WJN{K\log_{J}{m}}+(L_{\rm max}-WJ)N{K\log_{J}{m}})$.  Note that this complexity is an upper bound and an actual complexity becomes lower both when leaf nodes of the trees return duplicate codes simultaneously and when trees would have skewed structures.  Also, the search complexity for priority queue is negligible compared to the tree search complexity. In sum up, the overall complexity can be well-approximated as 
\begin{equation}\label{eq:appcomp}
\mathcal{O}(N L_{\rm max}({{K\log_{J}{m}}}+1)),
\end{equation} where $N L_{\rm max}$ accounts for the detection complexity in (\ref{lowcom}). It is noticeable that the complexity of the proposed low-complexity SL detector grows linear with $K$ while the other SL detectors grow exponentially with $K$. The approximated complexity in (\ref{eq:appcomp}) will be used in Section \ref{simulation}  to compute the complexity of the proposed LSL detector.

\section{Numerical results}\label{simulation}

We evaluate the average bit-error-rate (BER) performances of the proposed SL detector over the existing SL detectors in \cite{pohang}. Also, it is shown that the proposed low-complexity SL detector achieves the original performance with much lower complexity.   
For the simulations, QPSK modulation and Rayleigh fading are assumed.  
When a training overhead is small (e.g., $T$ is small), an empirical error-probability (e.g., $\epsilon_{c,i}$) can be underestimated as zero although it is indeed not. Since this can cause severe error-floor problem, we assign a minimum value of $\hat{\epsilon}_{c,i}$ as $10^{-3}$. 


\begin{figure}
\centerline{\includegraphics[width=9.5cm,height=7cm]{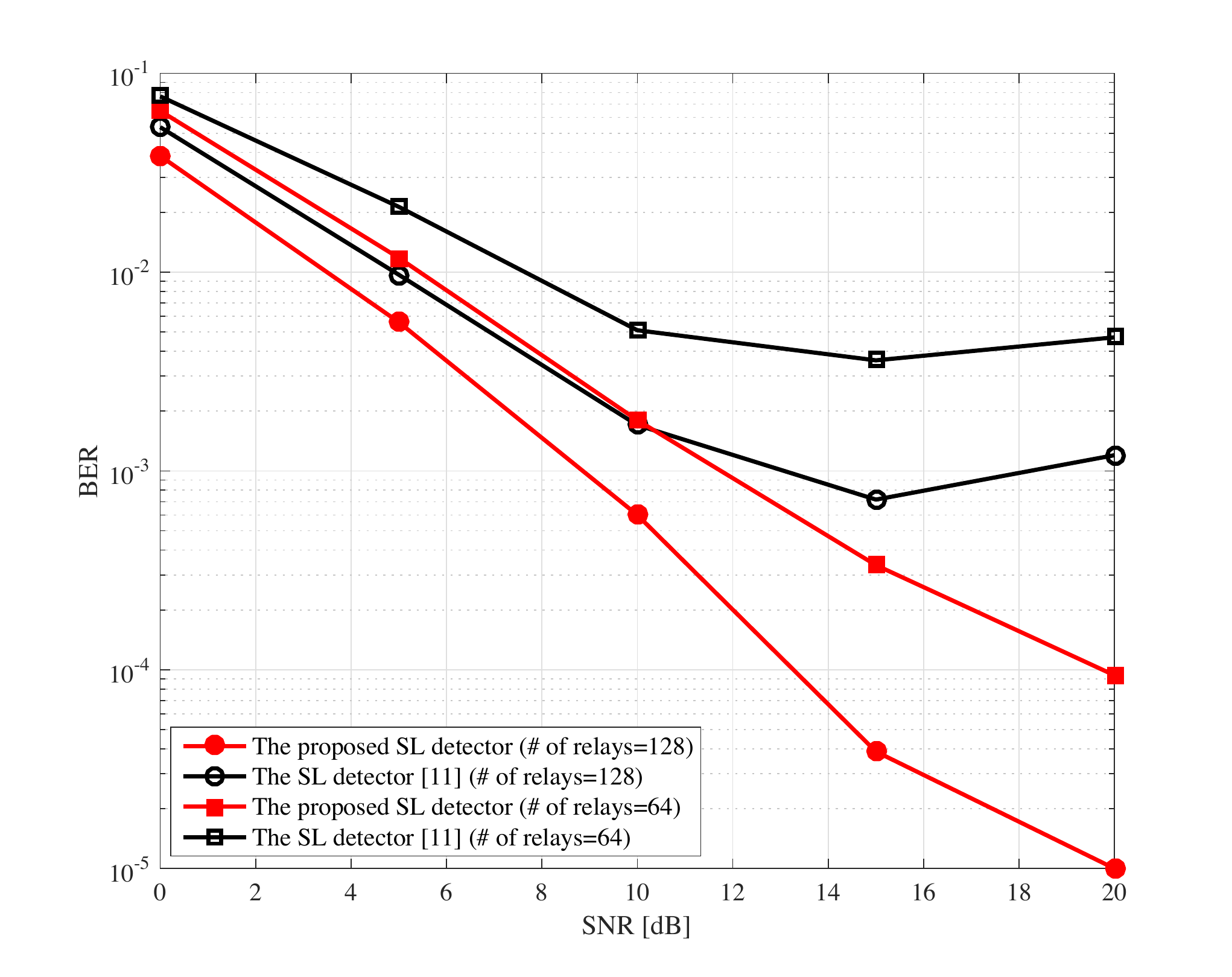}}
\caption{$K=8$, $N_{\rm r}=64$, and $T=15$. Performance comparisons of the proposed SL detector and the existing SL detectors.}
\label{comp1}
\end{figure}

Fig.~\ref{comp1} shows the BER performances of the proposed SL detector and the existing one in \cite{pohang}. Here, the number of training for each $c$ is set by 15 (e.g., $T=15$). We considered the two hop distributed reception network for the following two scenarios: i) 64 intermediate relays; ii) 128 intermediate relays. From Fig.~\ref{comp1}, we can observe that the proposed SL detector outperforms the existing SL detector, which implies that the proposed Bernoulli-like model is more suitable to binary data than Gaussian model.

\begin{figure}
\centerline{\includegraphics[width=9.5cm,height=7cm]{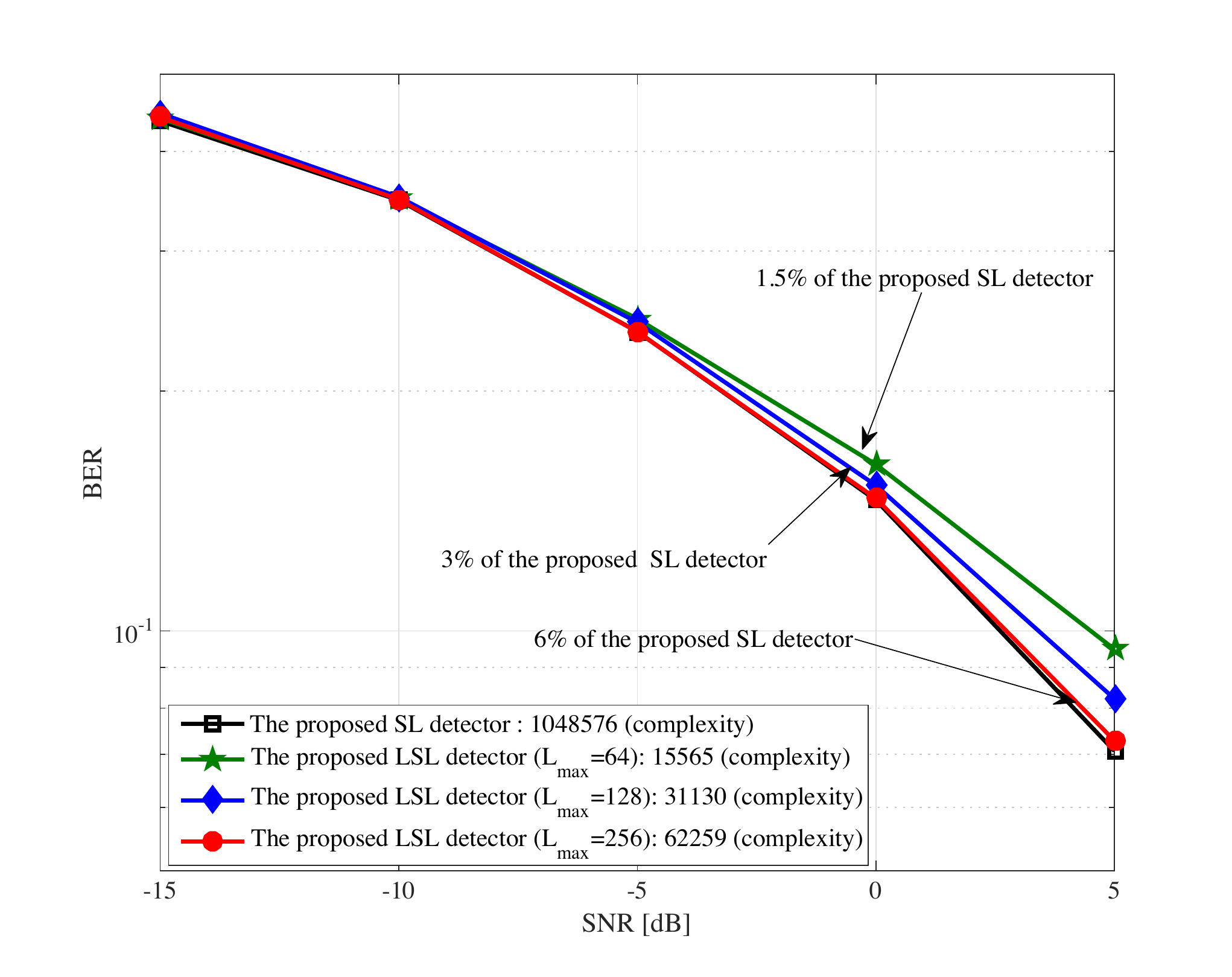}}
\caption{$K=14$, $N_{\rm r}=64$, and $T=15$. Performances of the proposed low-complexity SL detector according to the size of reduced search space.}
\label{comp3}
\end{figure}

Fig.~\ref{comp3} shows the BER performances of the proposed low-complexity detector according to $L_{\rm max}$ in Algorithm 2. Also, we set $J=32$ in Algorithm 1.  In this simulation, the benefit of low-complexity detector stands out, since it can achieve the optimal performance perfectly with only $6\%$ of original complexity. Thus, it is expected that the use of low-complexity technique is more beneficial for a large-scale distributed reception system (e.g., a large $K$).



\section{Conclusion}\label{conclusion}
We proposed a supervised-learning (SL) detector by introducing a novel  Bernoulli-like model as data probability distribution. Differently from a widely used Gaussian model, it can exploit the structure of binary data (e.g., one-bit quantized observation). We further developed the low-complexity SL detector with the aid of the  fast kNN algorithm. Simulation results demonstrated that the proposed low-complexity detector almost achieves the original performance with a significantly lower complexity. Therefore, the proposed detector would be a good practical candidate for multihop distributed reception systems. We would like to emphasize that the proposed SL detector can be straightforwardly applied to any multihop relay network with a single destination. On going work, we are investigating to generalize the Bernoulli-like model by capturing the correlation of elements in an observed data. Also, it is an interesting future work to extend the proposed SL detector for the multihop communication systems with low-resolution ADCs (e.g., 2$\sim$ 3-bit ADCs).

\end{document}